\documentclass[aps,prl,twocolumn,groupedaddress,showpacs]{revtex4}
\usepackage{graphicx}
\begin{document}

\title{Passive Sliders on Growing Surfaces and (anti-)Advection in Burger's Flows}

\author{Barbara Drossel}
\affiliation{Institut f\"ur Festk\"orperphysik, TU Darmstadt,
Hochschulstr.~6, 64289 Darmstadt, Germany }

\author{Mehran Kardar}
\affiliation{Department of Physics, Massachusetts
Institute of Technology, Cambridge, Massachusetts 02139}

\date{\today}

\begin{abstract}
We study the  fluctuations of particles sliding on a stochastically growing surface.
This problem can be mapped to motion of passive scalars in
a randomly stirred Burger's flow.
Renormalization group studies, simulations, and scaling arguments in one dimension, 
suggest a rich set of phenomena:
If  particles slide with the avalanche of growth sites (advection with the fluid), 
they tend to cluster and follow the surface dynamics.
However, for particles sliding against the avalanche (anti-advection),
we find slower diffusion dynamics, and density fluctuations with no simple relation to
the underlying fluid, possibly with continuously varying exponents.
\end{abstract}

\pacs{68.35.Rh, 02.50.Ey, 64.60.Ht}
%\keywords{}

\maketitle

The advection of a passive scalar, such as dye, a temperature field,
or dust particles, in a turbulent flow is a scientific problem of great interest,
since the fluctuations of the scalar field appear to have
no simple relation to the statistics of the flow \cite{siggiareview}. 
This holds not only for fully developped
turbulence, as emerging from the Navier-Stokes equation, but even for
the much simpler case of a random Gaussian velocity field which has
power-law correlations in space and is delta-correlated in time \cite{kraichnan}.

In this article, we introduce and study a related problem of particles passively
sliding on a fluctuating landscape, such as a growing surface.
To exploit its mathematical similarity to the better known problem of advection of
a passive scalar, we start by considering a velocity field
$ \vec v(\vec x,t)$ which follows the Burger's equation, given by
\begin{equation}
\frac{\partial \vec v}{\partial t} + \lambda(\vec v
\cdot \nabla) \vec v = \nu \nabla^2 \vec v + \nabla
\zeta_h(\vec x,t). \label{burgers}
\end{equation}
We also impose the additional condition 
$$ \vec v = -\nabla h, $$
which ensures a gradient flow with no vortices. 
This condition is preserved by the random stirring with the gradient 
of a Gaussian white noise $\zeta_h$, with correlations
\begin{displaymath}
%\label{white-noise}
\langle \zeta_h(\vec x,t) \zeta_h(\vec x\,',t')\rangle=2D_h 
\delta^d(\vec x - \vec x\,')\delta(t-t').
\end{displaymath}
The noise term also prevents the formation of shock waves.
Shock waves occur in the absence of a
stirring force, or with stirring correlated over large spatial scales \cite{mezard}. 
In contrast to the Navier-Stokes equation, Burger's equation has no pressure term, and
the fluid is therefore compressible. 
Our use of Burger's equation is not so much motivated by
describing a particular physical situation, but because its flows
are simpler than Navier-Stokes turbulence, which is a yet unsolved problem.
When expressed in terms of the scalar variable $h(\vec x,t)$, Eq.~(\ref{burgers}) 
becomes the Kardar-Parisi-Zhang (KPZ) equation \cite{KPZ} for interface growth,
\begin{equation}\label{kpz}
\frac{\partial h}{\partial t} = {\nu} \nabla^2 h + \frac{{\lambda}}{2}
(\nabla h)^2 + \zeta_h(\vec x,t),
\end{equation}
where $h(\vec x,t)$ is the height of interface at position $\vec x$ at time $t$. 
Fluctuations of the surface height obey the scaling relation \cite{FV85}
\begin{equation}\label{hcorr}
\langle \left[h(\vec x,t)-h(\vec x\,',t')\right]^2\rangle =
\left|\vec x-\vec x\,'\right|^{2\chi} g\left[\frac{t-t'}{\left|\vec x-\vec x\,'\right|^{z}}\right] .
\end{equation}
In one dimension, the only case that we shall consider here,
the roughness exponent is $\chi=1/2$, and the dynamic exponent is $z=3/2$. 
These values are exact due to a fluctuation-dissipation condition,
and the Galilei invariance of Eq.~(\ref{burgers}) \cite{fns}.

A particle moved by the flow is characterized by the Langevin-equation
\begin{equation} 
\frac{d \vec x}{dt} = a\vec v  + \zeta_x(t),
\label{langevin}
\end{equation}
where the white noise $\zeta_x$ describes the effects of the
temperature and has a correlation function
\begin{displaymath}
\langle \zeta_x(t)\zeta_x(t')\rangle =2\kappa 
%-\nabla^2\delta^d(\vec x-\vec x\,')
\delta(t-t')\, .
\end{displaymath} 
For $a=\lambda$, the particle is advected in the same way as an
element of the fluid. For $0<a<\lambda$, the particle is slower than
the neighboring fluid elements, while a negative $a$ describes a
particle moving against the flow ({\em anti-advection}). 
 In the language of the fluctuating interface (with $\vec v = -\nabla h$),
Eq.~(\ref{langevin}) describes a particle sitting on the surface and
sliding with a velocity proportional to the local slope, 
for instance under the influence of gravitation. 
(This equation also describes the dynamics of a single domain wall 
separating two phases \cite{dro00}.)
The corresponding equation for the particle density $\rho(\vec x,t)$ is
\begin{equation}
\frac{\partial \rho}{\partial t} = {\kappa} \nabla^2\rho+{a}\nabla
(\rho \nabla h)+\zeta_{\rho}(\vec x,t), \label{rho}
\end{equation}
where we have again added a white noise, with 
$$\langle \zeta_\rho(\vec x,t)\zeta_\rho(\vec x\,',t')\rangle =-2D_\rho
\nabla^2 \delta^d(\vec x-\vec x\,')\delta(t-t')\, .$$ 
In this paper, we consider only noninteracting particles that move independently of each
other. The case of many particles with hard-core repulsion is treated in Ref.~\cite{das01}.

The difficulty in analysis of the stochastic Eqs.~(\ref{kpz},\ref{rho}) arises
from the non-linear couplings. 
A perturbative treatment of the non-linearities results in divergences that
have to be controlled, e.g. by a renormalization group (RG) treatment.
Such a treatment is well-known for the KPZ equation~(\ref{kpz}), and
can be extended to the advection equation~(\ref{rho}).
In particular, at the one loop order, the reduced parameters 
$\alpha=a/\lambda$ and $k=\kappa/\nu$, satisfy the RG flows
\begin{eqnarray}
\frac{d {\alpha }}{d l}&=&{\alpha }\left[z_\rho-\frac{3}{2}+
\frac{2({\alpha }-{ \alpha }^2)}{(1+{k})^2}\right]\, ,\nonumber\\
\frac{d {k}}{dl}&=&{k}\left[z_\rho-2+\frac{{ \alpha }^2(3-{k})}{{k}(1+{k})^2}\right]\, ,\nonumber\\
\frac{d D_\rho}{dl}&=&D_\rho\left[z_\rho-2\chi_\rho-1+2\frac{{ \alpha }^2}{{k}(1+{k})}\right]\, .
\label{flow}
\end{eqnarray}
Similar to Eq.~(\ref{hcorr}), the exponents $\chi_\rho$ and $z_\rho$ describe 
the scaling of  particle density fluctuations via
\begin{equation}
\langle \left[\rho(x,t)-\rho(x\,',t')\right]^2\rangle \sim \left|x-x\,'\right|^{2\chi_\rho-2}
g\left[\frac{t-t'}{\left|\vec x-\vec x\,'\right|^{z_\rho}}\right] .
\label{chirho}
\end{equation}
Standard dynamic RG calls for a single exponent $z$ for the relative scaling of
time and space. The reason for including a distinct $z_\rho$ for the dynamics of
particles shall be explained below.
Other than this difference, the above RG equations are a special case
of the flows derived in Ref.~\cite{deniz} for a polymer drifting
through a random medium.  

Equations (\ref{flow}) show quite distinct behavior depending on the sign of $\alpha=a/\lambda$.
While $\alpha>0$ is the natural choice for advection by a fluid, both signs
are possible for particles sliding on a fluctuating surface.
Indeed, we first encountered $\alpha<0$ (anti-advection) in connection with domain
walls in a model of growing surfaces \cite{dro00}, inspiring the further simulations
described below.
For $\alpha>0$, the RG equations have an attractive fixed point at finite parameter
values, with the standard dynamic exponent $z=z_\rho=3/2$.
Using $z_\rho=3/2$, it is easy to show that the first two RG equations (\ref{flow}) 
yield the fixed point values $\alpha^*=1$ and the $k^*=1$, while the third sets $\chi_\rho=3/4$.
Note that for $\alpha=k=1$, the density fluctuations share the Galilei-invariance of the fluid flows,
and since symmetries are preserved under renormalization, 
this fixed point could have been predicted without any calculations. 
However, this symmetry does not fix the scaling dimension $\chi_\rho$ 
(in contrast to the values of $\chi$ and $z$),
and we shall in fact show below that the observed value of $\chi_\rho=1/2$ is  
different from the above one-loop prediction.
The case $a/\lambda>0$ is the one that is usually considered in related work, 
as for instance in Ref.~\cite{fns}, which contains a paragraph on passive scalar advection in
randomly stirred flows; in Ref.~\cite{DE99} which discusses (among others) a
situation that is equivalent to the advection of a single particle by
Burger's flow; in Ref.~\cite{CSC} which focusses on the world lines of particles,  
and in Ref.~\cite{das01}.

For $a/\lambda < 0$ and $z_\rho=z$, the RG flows drive the parameters 
$\alpha$ and $k$ towards zero, such that to leading order $\alpha^2/k=1/6$ 
(and $\chi_\rho=5/12$). 
The vanishing of the parameter $a$ under the coarse-graining action of RG
suggests that on larger scales the particles are less able to perform their sliding motion. 
A potential interpretation of the lack of a standard fixed point is that the particles
do not follow the fluctuating surface, and are instead governed by a slower timescale.
(A similar vanishing of parameters under RG occurs for decoupled equations
with distinct time scales.) 
Allowing for a distinct value of  $z_\rho$ in the physically meaningful interval 
$(3/2, 2)$, does indeed lead to nonzero fixed points for $\alpha$ and $k$.
Interestingly, the resulting exponent is non-universal, varying continuously
along the fixed line $\alpha^*(k)$.
This conclusion  appears to be supported by the 
simulations described below.
However, we shall make no quantitative claims based on this RG study,
both because of the unconventional use of two timescales, 
and the ``strong-coupling'' nature of the fixed point, which precludes controlled
perturbation expansion in any dimension.

Next, we present the results of computer simulations. 
The growing interface is obtained from a restricted solid-on solid model, by
adding ``bricks'' to a growing ``wall''.  Starting from a flat surface
and imposing periodic boundary conditions, bricks are added to
randomly chosen sites such that no overhangs are formed, and with the
center of each brick atop the intersection of two bricks in the layer below. 
Such a growth rule leads to surfaces that are in the same
universality class as the KPZ equation.
The parameter $\lambda$ is negative in this model as the average
growth velocity is less for a sloped surface.
The advected particles are intially placed at random surface positions. 
Whenever a brick is added to the site of a particle, it is moved to the {\em lower} one 
of its two nearest-neighbor sites with probability $(1+K)/2$, and to the {\em higher}
nearest-neighbor site with probability $(1-K)/2$.
The parameter $K$ is adjustable, but if not explicitly 
mentioned in the following, it is set to $K=1$. 
When both nearest-neighbor sites are at the same height, the particle is moved
to the right or left with equal probability. 
The downhill character of these moves corresponds to $a>0$ in Eq.~(\ref{langevin}).  
The case $a<0$ is implemented similarly, moving particles uphill instead of downhill. 

\begin{figure}
\includegraphics*[width=5.99cm]{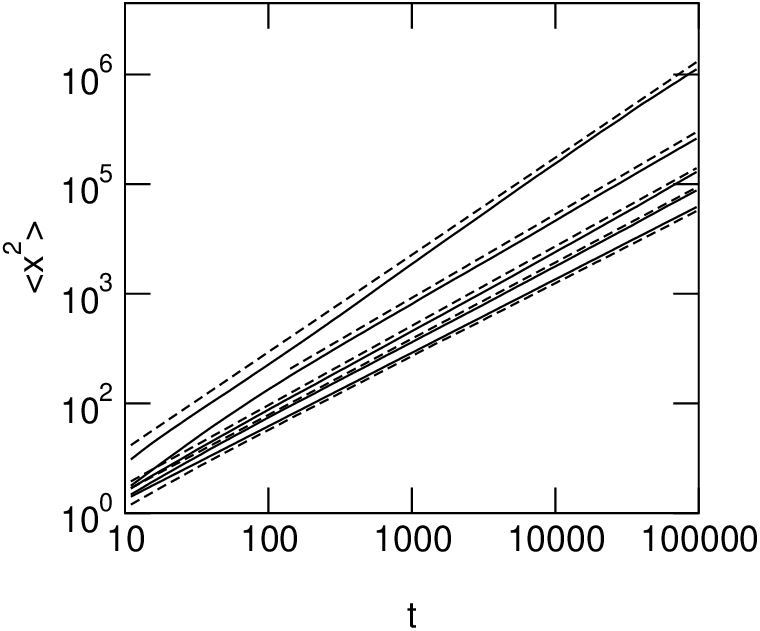}
\caption{The mean square distance travelled by the sliding particles as
function of time. The top curve is for $a/\lambda>0$ (advection), while the others are
for $a/\lambda<0$ (anti-advection) with $K=1$, 0.25, 0.125, 0.025 (from top to
bottom).  The dashed straight lines are power-law fits to
exponents $2/z_\rho$, with $z_\rho$ equal to 1.5, 1.74, 1.83, 1.91,
1.98 from top to bottom. \label{fig1}}
\end{figure}

After a time period long enough for establishing a stationary state, the
diffusion of the particles and the particle densities were quantified. 
Figure~\ref{fig1} shows the mean square distance 
$\langle x^2(t)\rangle \sim t^{2/z_\rho}$ travelled by a particle as function of time, for
different values of the simulation parameters.
As suggested by the RG analysis, the particles can follow the surface fluctuations 
for $a/\lambda>0$, and  their diffusion is characterized by the same dynamical exponent
$z_\rho=3/2$ as the surface. 
For $a/\lambda<0$ (particles moving downhill), the diffusion exponent
$z_\rho$ is larger, and it seems to be nonuniversal, i.e., it
depends on the value of the parameter $K$.

\begin{figure}
\includegraphics*[width=5.99cm]{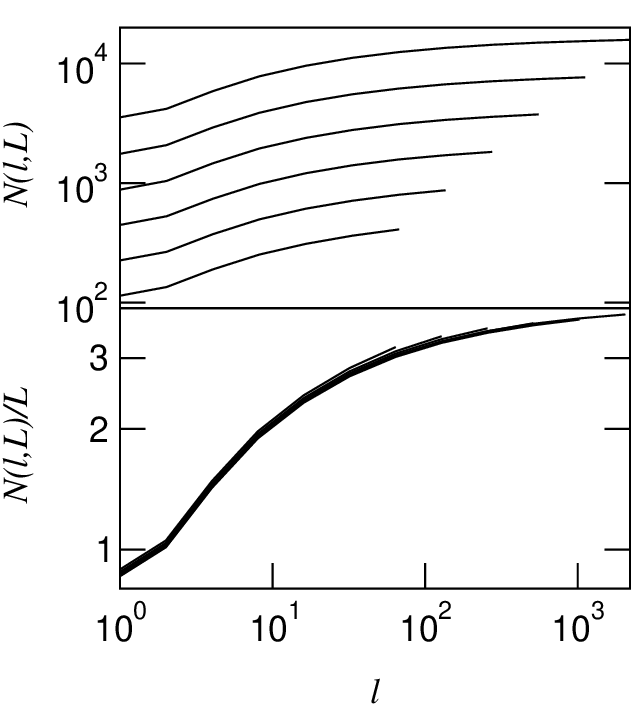}
\caption{Mean number of particles in the same bin of size $l$ as 
a randomly selected particle, for $a/\lambda>0$ and different $L$.
\label{fig2}}
\end{figure}

This qualitative difference between the cases $a/\lambda>0$ and
$a/\lambda<0$ is also visible in the correlations of particle density.
To examine this quantity, we divided the system into $L/l$ bins of size $l$
and counted the number of particles in each bin. 
Let us choose a particle at random, and denote by $N(l,L)$ the number
of particles that are in the same bin.
This quantity is plotted in Fig.~\ref{fig2} as function of $l$, for $a/\lambda>0$. 
Using the scaling form of Eq.~(\ref{chirho}), we have
\begin{equation}
\label{NlL}
N(l,L)\sim \left\langle \int_{-l/2}^{l/2} dx \rho(x,t)\rho(0,t)dx\right\rangle \sim
l^{2\chi_\rho-1}L^{2-2\chi_\rho},
\end{equation}
where the dependence on $L$ follows from $N(L,L)$ being the total number 
of particles in the system, which was chosen proportional to the system size $L$. 
The curves in Fig.~\ref{fig2} are in fact proportional to $L$, and become
constant for larger $l$, both results consistent with $\chi_\rho=1/2$ in Eq.~(\ref{NlL}).
This value is different from the 1-loop RG result obtained above.
\begin{figure}
\includegraphics*[width=5.99cm]{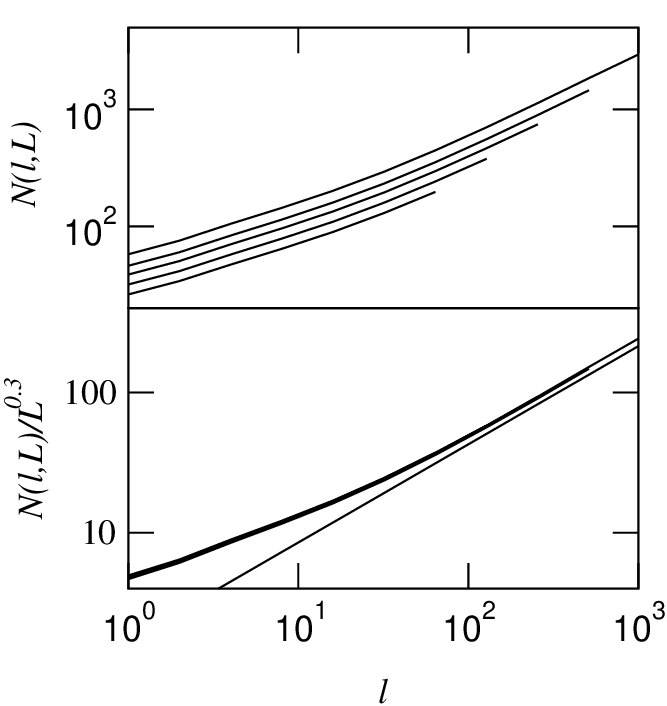}
\caption{Mean number of particles in the same bin of size $l$ as a 
randomly selected particle, for $a/\lambda<0$ and different $L$.
The straight line in the bottom figure is a power law with exponent 0.7.
\label{fig3}}
\end{figure}
The corresponding results for $a/\lambda < 0$ are shown in Fig.~\ref{fig3} (for $K=1$). 
Since particles cannot fully follow the surface fluctuations, the
proportion of particles within a given interval around a randomly
chosen particle increases with increasing system size, and we measured
a nontrivial scaling exponent $\chi_\rho\simeq 0.85$. 
For smaller values of $K$, the scaling behavior is not so clear. 
The asymptotic scaling of the factor that maps curves for different $L$ 
onto each other could not be determined reliably. 
It appears, however, that $\chi_\rho$ is nonuniversal, becoming larger with 
decreasing $K$ (and increasing particle diffusion exponent $z_\rho$). 

Particle clustering, which is the most prominent feature of our
system, does not occur in Navier-Stokes turbulence, where
the fluid is incompressible. Instead, fluid elements are streched and
folded, leading ultimately to a homogeneous particle distribution,
unless new particles are continuously added.  

\begin{figure}
\includegraphics*[width=5.99cm]{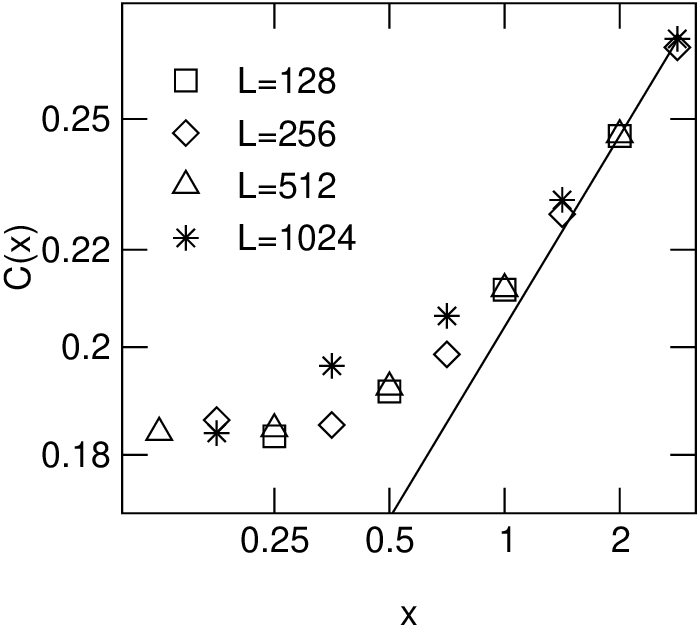}
\caption{The scaling function for the particle velocity as a function of
slope and system size, for $a/\lambda< 0$ and $K=1$. 
The straight line has the asymptotic slope $2y$, with $y=0.14$. 
\label{fig4}}
\end{figure}

In order to better understand the nonuniversal and nontrivial scaling
for $a/\lambda<0$, we measured directly the velocity of particles as a
function of an external slope. A particular slope was
imposed on the system by choosing periodic boundary conditions with a
fixed height difference between the two ends. 
The average speed of particles was then evaluated as function of slope and
system size. As expected, for $a/\lambda>0$ the mean velocity is
simply proportional to the slope. For $a/\lambda<0$, however, the
velocity shows a nontrivial scaling behavior of the form
\begin{equation}
v = L^{-y} \nabla h\; {\cal C}(\nabla h \sqrt{L}),
\label{vscaling}
\end{equation}
with $y\simeq 0.14$, as shown in Figure \ref{fig4}. 
The scaling function ${\cal C}(x)$ is constant for small $x$ and increases as
$x^{2y}$ for large $x$, such that $v\propto (\nabla h)^{1+2y}$ is independent of $L$.
(Note that the data for $L=1024$ show large fluctuations and seem
sensitive to the choice of the random number generator, and therefore
not all data points collapse with those for smaller $L$.)  Just as for
the density correlations, the data for smaller values of $K$ did not
show nice scaling behavior for the system sizes in our simulations.
Equation~(\ref{vscaling}) implies that the factor $a$ relating
velocity to slope [see Eq.~(\ref{langevin})] decreases as a function
of system size, consistent with the RG prediction.  However, while the
1-loop result suggests an inverse logarithmic dependence of $a$ on the
length scale (implying $y=0$), the value of $y$ found in the
simulations is nonzero.

The form of Eq.~(\ref{vscaling}) can be justified by the following argument: 
The growth rules for the surface imply that sequences of brick addition 
usually move from local minima in the uphill direction. 
The reason is that the addition of a brick generates a potential growth site 
(where a brick can be added without generating overhangs) 
at the nearest uphill position. 
The particles that try to slide downhill are therefore faced with an upward 
avalanche of growth mounds of different sizes that hamper their downhill motion. 
(This is analogous to a moving fluid density fluctuation faced by a
particle advecting against the flow.)
The horizontal size of the largest mounds, $l$, scales as 
$L\tilde{\mathcal{C}}(\nabla h \sqrt{L})$, with $\tilde{\mathcal{C}}(x)$ being
constant for small $x$ and $\tilde{\mathcal{C}}(x) \sim x^{-2}$ for large $x$. 
If we assume that the particle velocity is slowed down by some power of the 
size of the largest mounds, we obtain Eq.~(\ref{vscaling}).
The exponent $y$ may then have a non-universal dependence on
the manner in which the particle goes over these barriers.

Finally, we shall try to compute the exponent $y$ by relating characteristic
time scales of surface and particle fluctuations.
Consider a portion of the surface of size $l$, with a typical slope of order 
$|\nabla h|\sim l^{-1/2}$.
For this slope to change by surface fluctuations, we should wait a time
of the order $t\sim l^z\sim l^{3/2}$.
During this time the advecting particle will move against the slope by
an amount 
$$x \sim {a(l)}l^{-1/2}t \sim t^{2(1-y)/3}\, ,$$
leading to the scaling relation
\begin{equation}
z_\rho =\frac{3}{2(1-y)}\, .\label{yscaling}
\end{equation}
For $y\simeq 0.14$, this gives $z_\rho \simeq 1.74$, which is the
value measured in simulations of particle diffusion for $K=1$. 

In conclusion, we have found surprisingly intricate scaling phenomena for
passive sliders on stochastically growing surfaces, and the equivalent
problem of passive scalar motion in Burger's turbulence.
In particular, there is a qualitative difference between   the cases
$a/\lambda>0$ and $a/\lambda<0$, describing
respectively passive advection with or against the flow. 
The former leads to clustering, while the latter yields non-trivial exponents.
This conclusion is bolstered by RG analysis, simulations, and scaling arguments. 
While the simulations provide convincing indication of a
nonuniversal value of the exponent $z_\rho$ characterizing particle
diffusion(for $a/\lambda<0$), the evidence is less compelling for
as the density correlations.
The absence of nice scaling for most  values of simulation parameters
may indicate slow crossovers, but then also leaves open the possibility that the 
exponent $z_\rho$ is universal even in the case $a/\lambda<0$.

\begin{acknowledgments}
We thank Kay Wiese for helpful discussions. 
B.D. was supported by the Deutsche 
Forschungsgemeinschaft (DFG) under Contract No Dr300-2/1.
M.K. is supported by the NSF through grant number DMR-01-18213.
\end{acknowledgments}

\end{document}